

\newcount\mgnf  
\mgnf=0 

\ifnum\mgnf=0
\def\openone{\leavevmode\hbox{\ninerm 1\kern-3.3pt\tenrm1}}%
\def\*{\vglue0.3truecm}\fi
\ifnum\mgnf=1
\def\openone{\leavevmode\hbox{\ninerm 1\kern-3.63pt\tenrm1}}%
\def\*{\vglue0.5truecm}\fi

%
%
%
%
%
%
%

\global\newcount\numsec\global\newcount\numapp
\global\newcount\numfor\global\newcount\numfig
\global\newcount\numsub
\numsec=0\numapp=0\numfig=0
\def\veroparagrafo{\number\numsec}\def\veraformula{\number\numfor}
\def\veraappendice{\number\numapp}\def\verasub{\number\numsub}
\def\verafigura{\number\numfig}
\def\section(#1,#2){\advance\numsec by 1\numfor=1\numsub=1\numfig=1%
\SIA p,#1,{\veroparagrafo} %
\write15{\string\Fp (#1){\secc(#1)}}%
\write16{ sec. #1 ==> \secc(#1)  }%
\hbox to \hsize{\titolo\hfill \number\numsec. #2\hfill%
\expandafter{\alato(sec. #1)}}\*}

\def\appendix(#1,#2){\advance\numapp by 1\numfor=1\numsub=1\numfig=1%
\SIA p,#1,{A\veraappendice} %
\write15{\string\Fp (#1){\secc(#1)}}%
\write16{ app. #1 ==> \secc(#1)  }%
\hbox to \hsize{\titolo\hfill Appendix A\number\numapp. #2\hfill%
\expandafter{\alato(app. #1)}}\*}

\def\senondefinito#1{\expandafter\ifx\csname#1\endcsname\relax}

\def\SIA #1,#2,#3 {\senondefinito{#1#2}%
\expandafter\xdef\csname #1#2\endcsname{#3}\else
\write16{???? ma #1#2 e' gia' stato definito !!!!} \fi}

\def \Fe(#1)#2{\SIA fe,#1,#2 }
\def \Fp(#1)#2{\SIA fp,#1,#2 }
\def \Fg(#1)#2{\SIA fg,#1,#2 }

\def\etichetta(#1){(\veroparagrafo.\veraformula)%
\SIA e,#1,(\veroparagrafo.\veraformula) %
\global\advance\numfor by 1%
\write15{\string\Fe (#1){\equ(#1)}}%
\write16{ EQ #1 ==> \equ(#1)  }}

\def\etichettaa(#1){(A\veraappendice.\veraformula)%
\SIA e,#1,(A\veraappendice.\veraformula) %
\global\advance\numfor by 1%
\write15{\string\Fe (#1){\equ(#1)}}%
\write16{ EQ #1 ==> \equ(#1) }}

\def\getichetta(#1){\veroparagrafo.\verafigura%
\SIA g,#1,{\veroparagrafo.\verafigura} %
\global\advance\numfig by 1%
\write15{\string\Fg (#1){\graf(#1)}}%
\write16{ Fig. #1 ==> \graf(#1) }}

\def\etichettap(#1){\veroparagrafo.\verasub%
\SIA p,#1,{\veroparagrafo.\verasub} %
\global\advance\numsub by 1%
\write15{\string\Fp (#1){\secc(#1)}}%
\write16{ par #1 ==> \secc(#1)  }}

\def\etichettapa(#1){A\veraappendice.\verasub%
\SIA p,#1,{A\veraappendice.\verasub} %
\global\advance\numsub by 1%
\write15{\string\Fp (#1){\secc(#1)}}%
\write16{ par #1 ==> \secc(#1)  }}

\def\Eq(#1){\eqno{\etichetta(#1)\alato(#1)}}
\def\eq(#1){\etichetta(#1)\alato(#1)}
\def\Eqa(#1){\eqno{\etichettaa(#1)\alato(#1)}}
\def\eqa(#1){\etichettaa(#1)\alato(#1)}
\def\eqg(#1){\getichetta(#1)\alato(fig. #1)}
\def\sub(#1){\0\palato(p. #1){\bf \etichettap(#1).}}
\def\asub(#1){\0\palato(p. #1){\bf \etichettapa(#1).}}

\def\equv(#1){\senondefinito{fe#1}$\clubsuit$#1%
\write16{eq. #1 non e' (ancora) definita}%
\else\csname fe#1\endcsname\fi}
\def\grafv(#1){\senondefinito{fg#1}$\clubsuit$#1%
\write16{fig. #1 non e' (ancora) definito}%
\else\csname fg#1\endcsname\fi}
\def\secv(#1){\senondefinito{fp#1}$\clubsuit$#1%
\write16{par. #1 non e' (ancora) definito}%
\else\csname fp#1\endcsname\fi}

\def\equ(#1){\senondefinito{e#1}\equv(#1)\else\csname e#1\endcsname\fi}
\def\graf(#1){\senondefinito{g#1}\grafv(#1)\else\csname g#1\endcsname\fi}
\def\figura(#1){{\css Figura} \getichetta(#1)}
\def\secc(#1){\senondefinito{p#1}\secv(#1)\else\csname p#1\endcsname\fi}
\def\sec(#1){{\secc(#1)}}
\def\refe(#1){{[\secc(#1)]}}

\def\BOZZA{\bz=1
\def\alato(##1){\rlap{\kern-\hsize\kern-1.2truecm{$\scriptstyle##1$}}}
\def\palato(##1){\rlap{\kern-1.2truecm{$\scriptstyle##1$}}}
}

\def\alato(#1){}
\def\galato(#1){}
\def\palato(#1){}


{\count255=\time\divide\count255 by 60 \xdef\hourmin{\number\count255}
        \multiply\count255 by-60\advance\count255 by\time
   \xdef\hourmin{\hourmin:\ifnum\count255<10 0\fi\the\count255}}

\def\oramin{\hourmin }

\def\data{\number\day/\ifcase\month\or gennaio \or febbraio \or marzo \or
aprile \or maggio \or giugno \or luglio \or agosto \or settembre
\or ottobre \or novembre \or dicembre \fi/\number\year;\ \oramin}
\setbox200\hbox{$\scriptscriptstyle \data $}

\newdimen\xshift \newdimen\xwidth \newdimen\yshift \newdimen\ywidth

\def\ins#1#2#3{\vbox to0pt{\kern-#2\hbox{\kern#1 #3}\vss}\nointerlineskip}

\def\eqfig#1#2#3#4#5{
\par\xwidth=#1 \xshift=\hsize \advance\xshift
by-\xwidth \divide\xshift by 2
\yshift=#2 \divide\yshift by 2
\line{\hglue\xshift \vbox to #2{\vfil
#3 \includegraphics{#4.ps}
}\hfill\raise\yshift\hbox{#5}}}

\def\8{\write12}  


\let\a=\alpha \let\b=\beta  \let\g=\gamma   \let\e=\varepsilon
  \let\h=\eta   \let\th=\theta \let\k=\kappa \let\l=\lambda
    \let\n=\nu         \let\p=\pi    \let\r=\rho
\let\s=\sigma \let\t=\tau    
   \let\o=\omega
\let\G=\Gamma \let\D=\Delta  \let\Th=\Theta\let\L=\Lambda

\def\\{\hfill\break} \let\==\equiv

\let\io=\infty 
\def\ap{{\it a priori\ }}
\let\0=\noindent

\def\ie{{i.e. }}\def\eg{{e.g. }}
\let\dpr=\partial

\def\tende#1{\,\vtop{\ialign{##\crcr\rightarrowfill\crcr
 \noalign{\kern-1pt\nointerlineskip} \hskip3.pt${\scriptstyle
 #1}$\hskip3.pt\crcr}}\,}
\def\circage{\lower2pt\hbox{$\,\buildrel > \over {\scriptstyle \sim}\,$}}
\def\otto{\,{\kern-1.truept\leftarrow\kern-5.truept\to\kern-1.truept}\,}
\def\fra#1#2{{#1\over#2}}

\def\EE{{\cal E}}\def\MM{{\cal M}} \def\VV{{\cal V}}
 
 \def\BBB{{\cal B}}
\def\RR{{\cal R}}\def\LL{{\cal L}}  
\def\DD{{\cal D}} \def\SS{{\cal S}}

\def\T#1{{#1_{\kern-3pt\lower7pt\hbox{$\widetilde{}$}}\kern3pt}}
\def\VVV#1{{\underline #1}_{\kern-3pt
\lower7pt\hbox{$\widetilde{}$}}\kern3pt\,}
\def\W#1{#1_{\kern-3pt\lower7.5pt\hbox{$\widetilde{}$}}\kern2pt\,}
\def\Re{{\rm Re}\,}\def\Im{{\rm Im}\,}
\def\lis{\overline}\def\tto{\Rightarrow}

\def\indica{\leaders \hbox to 0.5cm{\hss.\hss}\hfill}
\def\guida{\leaders\hbox to 1em{\hss.\hss}\hfill}

\def\hh{{\bf h}}

\def\II{{\bf I}}

\def\ul{\underline}

\def\wt{\widetilde}


\def\to{\rightarrow}

\let\ciao=\bye
\def\qed{\hfill\raise1pt\hbox{\vrule height5pt width5pt depth0pt}}
\def\Val{{\rm Val}}

\def\RRR{\hbox{\msytw R}} 
 \def\CCC{\hbox{\msytw C}}

 \def\ZZZ{\hbox{\msytw Z}}
 
\def\TTT{\hbox{\msytw T}}

\def\ul#1{{\underline#1}}

\def\V0{{\bf 0}}
\def\defi{\,{\buildrel def\over=}\,}


\newcount\tipobib\newcount\bz\bz=0\newcount\aux\aux=1
\newdimen\bibskip\newdimen\maxit\maxit=0pt


\tipobib=2
\def\9#1{\ifnum\aux=1#1\else\relax\fi}

\newwrite\bib
\immediate\openout\bib=\jobname.bib
\global\newcount\bibex
\bibex=0
\def\verabib{\number\bibex}

\ifnum\tipobib=0
\def\cita#1{\expandafter\ifx\csname c#1\endcsname\relax
\hbox{$\clubsuit$}#1\write16{Manca #1 !}%
\else\csname c#1\endcsname\fi}
\def\rife#1#2#3{\immediate\write\bib{\string\raf{#2}{#3}{#1}}
\immediate\write15{\string\C(#1){[#2]}}
\setbox199=\hbox{#2}\ifnum\maxit < \wd199 \maxit=\wd199\fi}
\fi
\ifnum\tipobib=1
\def\cita#1{%
\expandafter\ifx\csname d#1\endcsname\relax%
\expandafter\ifx\csname c#1\endcsname\relax%
\hbox{$\clubsuit$}#1\write16{Manca #1 !}%
\else\probib(ref. numero )(#1)%
\csname c#1\endcsname%
\fi\else\csname d#1\endcsname\fi}%
\def\rife#1#2#3{\immediate\write15{\string\Cp(#1){%
\string\immediate\string\write\string\bib{\string\string\string\raf%
{\string\verabib}{#3}{#1}}%
\string\Cn(#1){[\string\verabib]}%
\string\CCc(#1)%
}}}%
\fi
\ifnum\tipobib=2%
\def\cita#1{\expandafter\ifx\csname c#1\endcsname\relax
\hbox{$\clubsuit$}#1\write16{Manca #1 !}%
\else\csname c#1\endcsname\fi}
\def\rife#1#2#3{\immediate\write\bib{\string\raf{#1}{#3}{#2}}
\immediate\write15{\string\C(#1){[#1]}}
\setbox199=\hbox{#2}\ifnum\maxit < \wd199 \maxit=\wd199\fi}
\fi

\def\Cn(#1)#2{\expandafter\xdef\csname d#1\endcsname{#2}}
\def\CCc(#1){\csname d#1\endcsname}
\def\probib(#1)(#2){\global\advance\bibex+1%
\9{\immediate\write16{#1\verabib => #2}}%
}

\def\C(#1)#2{\SIA c,#1,{#2}}
\def\Cp(#1)#2{\SIAnx c,#1,{#2}}

\def\SIAnx #1,#2,#3 {\senondefinito{#1#2}%
\expandafter\def\csname#1#2\endcsname{#3}\else%
\write16{???? ma #1,#2 e' gia' stato definito !!!!}\fi}

\bibskip=10truept
\def\hboxto{\hbox to}

\catcode`\{=12\catcode`\}=12
\catcode`\<=1\catcode`\>=2
\immediate\write\bib<
        \string\halign{\string\hboxto \string\maxit%
        {\string #\string\hfill}&%
        \string\vtop{\string\parindent=0pt\string\advance\string\hsize%
        by -.5truecm%
        \string#\string\vskip \bibskip
        }\string\cr%
>
\catcode`\{=1\catcode`\}=2
\catcode`\<=12\catcode`\>=12

\def\raf#1#2#3{\ifnum \bz=0 [#1]&#2 \cr\else
\llap{${}_{\rm #3}$}[#1]&#2\cr\fi}

\newread\bibin

\catcode`\{=12\catcode`\}=12
\catcode`\<=1\catcode`\>=2
\def\chiudibib<
\catcode`\{=12\catcode`\}=12
\catcode`\<=1\catcode`\>=2
\immediate\write\bib<}>
\catcode`\{=1\catcode`\}=2
\catcode`\<=12\catcode`\>=12
>
\catcode`\{=1\catcode`\}=2
\catcode`\<=12\catcode`\>=12

\def\makebiblio{
\ifnum\tipobib=0
\advance \maxit by 10pt
\else
\maxit=1.truecm
\fi
\chiudibib
\immediate \closeout\bib
\openin\bibin=\jobname.bib
\ifeof\bibin\relax\else
\raggedbottom
\input \jobname.bib
\fi
}

\openin13=#1.aux \ifeof13 \relax \else
\input #1.aux \closein13\fi
\openin14=\jobname.aux \ifeof14 \relax \else
\input \jobname.aux \closein14 \fi
\immediate\openout15=\jobname.aux

\def\biblio{\*\*\centerline{\titolo References}\*\nobreak\makebiblio}


\ifnum\mgnf=0
   \magnification=\magstep0
   \hsize=15truecm\vsize=20.8truecm\voffset2.truecm\hoffset.5truecm
   \parindent=0.3cm\baselineskip=0.45cm\fi
\ifnum\mgnf=1
   \magnification=\magstep1\hoffset=0.truecm
   \hsize=15truecm\vsize=20.8truecm
   \baselineskip=18truept plus0.1pt minus0.1pt \parindent=0.9truecm
   \lineskip=0.5truecm\lineskiplimit=0.1pt      \parskip=0.1pt plus1pt\fi


\mgnf=0   
\openin14=\jobname.aux \ifeof14 \relax \else
\input \jobname.aux \closein14 \fi
\openout15=\jobname.aux

\footline={\rlap{\hbox{\copy200}}\tenrm\hss \number\pageno\hss}
\def\fiat{}


\font\tenmib=cmmib10 
\font\sevenmib=cmmib7\font\fivemib=cmmib5 
\font\ottoit=cmti8
\font\fivei=cmmi5\font\sixi=cmmi6\font\ottoi=cmmi8
\font\ottorm=cmr8\font\fiverm=cmr5\font\sixrm=cmr6
\font\ottosy=cmsy8\font\sixsy=cmsy6\font\fivesy=cmsy5
\font\ottobf=cmbx8\font\sixbf=cmbx6\font\fivebf=cmbx5%
\font\ottott=cmtt8%
\font\ottocss=cmcsc8%
\font\ottosl=cmsl8%
\font\titolo=cmbx12
\font\titolone=cmbx10 scaled\magstep 2

\font\cs=cmcsc10
\font\sc=cmcsc10
\font\css=cmcsc8

\font\ninerm=cmr9
\font\ottorm=cmr8

\font\msytw=msbm9 scaled\magstep1


\def\ottopunti{\def\rm{\fam0\ottorm}%
\textfont0=\ottorm\scriptfont0=\sixrm\scriptscriptfont0=\fiverm%
\textfont1=\ottoi\scriptfont1=\sixi\scriptscriptfont1=\fivei%
\textfont2=\ottosy\scriptfont2=\sixsy\scriptscriptfont2=\fivesy%
\textfont3=\tenex\scriptfont3=\tenex\scriptscriptfont3=\tenex%
\textfont4=\ottocss\scriptfont4=\sc\scriptscriptfont4=\sc%
\textfont5=\tenmib\scriptfont5=\sevenmib\scriptscriptfont5=\fivei
\textfont\itfam=\ottoit\def\it{\fam\itfam\ottoit}%
\textfont\slfam=\ottosl\def\sl{\fam\slfam\ottosl}%
\textfont\ttfam=\ottott\def\tt{\fam\ttfam\ottott}%
\textfont\bffam=\ottobf\scriptfont\bffam=\sixbf%
\scriptscriptfont\bffam=\fivebf\def\bf{\fam\bffam\ottobf}%
\setbox\strutbox=\hbox{\vrule height7pt depth2pt width0pt}%
\normalbaselineskip=9pt\let\sc=\sixrm\normalbaselines\rm}
\let\nota=\ottopunti%

\textfont5=\tenmib\scriptfont5=\sevenmib\scriptscriptfont5=\fivemib
\mathchardef\Ba   = "050B  
\mathchardef\Bb   = "050C  
\mathchardef\Bg   = "050D  
\mathchardef\Bd   = "050E  
\mathchardef\Be   = "0522  
\mathchardef\Bee  = "050F  
\mathchardef\Bz   = "0510  
\mathchardef\Bh   = "0511  
\mathchardef\Bthh = "0512  
\mathchardef\Bth  = "0523  
\mathchardef\Bi   = "0513  
\mathchardef\Bk   = "0514  
\mathchardef\Bl   = "0515  
\mathchardef\Bm   = "0516  
\mathchardef\Bn   = "0517  
\mathchardef\Bx   = "0518  
\mathchardef\Bom  = "0530  
\mathchardef\Bp   = "0519  
\mathchardef\Br   = "0525  
\mathchardef\Bro  = "051A  
\mathchardef\Bs   = "051B  
\mathchardef\Bsi  = "0526  
\mathchardef\Bt   = "051C  
\mathchardef\Bu   = "051D  
\mathchardef\Bf   = "0527  
\mathchardef\Bff  = "051E  
\mathchardef\Bch  = "051F  
\mathchardef\Bps  = "0520  
\mathchardef\Bo   = "0521  
\mathchardef\Bome = "0524  
\mathchardef\BG   = "0500  
\mathchardef\BD   = "0501  
\mathchardef\BTh  = "0502  
\mathchardef\BL   = "0503  
\mathchardef\BX   = "0504  
\mathchardef\BP   = "0505  
\mathchardef\BS   = "0506  
\mathchardef\BU   = "0507  
\mathchardef\BF   = "0508  
\mathchardef\BPs  = "0509  
\mathchardef\BO   = "050A  
\mathchardef\BDpr = "0540  
\mathchardef\Bstl = "053F  

%
\fiat

\def\V#1{{\bf#1}}
\let\ig=\int


\centerline{\titolone Borel summability and Lindstedt series}

\vskip1.truecm 
\centerline{\titolo
O. Costin${}^{1}$,
G. Gallavotti${}^{2}$,
G. Gentile${}^{3}$ {\rm and} A. Giuliani${}^{4}$}
\vskip.2truecm
\centerline{{}$^{1}$ Department of Mathematics, The Ohio State University,
Columbus, Ohio 43210 USA}
\vskip.1truecm
\centerline{{}$^{2}$ Dipartimento di Fisica,
Universit\`a di Roma ``La Sapienza'', Roma I-00185 Italy}
\vskip.1truecm
\centerline{{}$^{3}$ Dipartimento di Matematica,
Universit\`a di Roma Tre, Roma I-00146 Italy}
\vskip.1truecm
\centerline{{}$^{4}$ Department of Physics, Princeton University, 
Princeton, New Jersey 08544 USA}
\vskip1.truecm

\line{\vtop{ \line{\hskip1truecm\vbox{\advance \hsize by -2.1 truecm
        \0{\cs Abstract.}  {\it Resonant motions of integrable systems
        subject to perturbations may continue to exist and to cover
        surfaces with parametric equations admitting a formal power
        expansion in the strength of the perturbation. Such series may
        be, sometimes, summed via suitable sum rules defining $C^\io$
        functions of the perturbation strength: here we find sufficient
        conditions for the Borel summability of their sums in the
        case of two-dimensional rotation vectors with Diophantine
        exponent $\t=1$ (\eg with ratio of the two independent
        frequencies equal to the golden mean).}} \hfill} }}

\*\*\*\*
\section(1,Introduction)

\0In the paradigmatic setting of KAM theory, one considers
unperturbed motions $\Bf\to\Bf+\Bo_0(\II)t$
on the torus $\TTT^d$, $d\ge 2$, driven by a Hamiltonian $H=H_0(\II)$,
where $\II\in\RRR^d$ are the actions conjugated to $\Bf$ and $\Bo_0(\II)=
\dpr_\II H_{0}(\II)$. Standard analytic KAM theorem
considers the perturbed Hamiltonian $H_\e=H_0(\II)+\e 
f(\Bf,\II)$, with $f$ analytic, and a frequency vector $\Bo_0$ which
is Diophantine with constants $C_0$ and $\t$ 
(\ie $|\Bo_0\cdot\Bn|\ge C_0|\Bn|^{-\t}$ $\forall\Bn\in\ZZZ^d$, 
$\Bn\neq\V0$) and which is among the frequencies of the
unperturbed system: $\Bo_0=\o_{0}(\II_0)$ for some $\II_0$.
Suppose $H_{0}(\II)=\II^{2}/2$ and $f(\Bf,\II)=f(\Bf)$ for simplicity,
that is assume that $H_{0}$ is quadratic and the perturbation
depends only on the angle variables. Then for $\e$ small enough
the unperturbed motion $\Bf\to\Bf+\Bo_0t$ can be analytically continued
into a motion of the perturbed system, in the sense that
there is an $\e$--analytic function $\hh_{\e}:\TTT^d\to\TTT^d$,
reducing to the  identity as $\e\to 0$, such that $\Bps+\hh_{\e}(\Bps)$
solves the Hamilton equations for $H_\e$, \ie
$\ddot \Bf + \e \dpr_{\Bf}f(\Bf)=\V0$, if $\Bps$ is replaced by
$\Bps+\Bo t$,  {\it for any choice of the
initial data $\Bps$}. The function $\hh_{\e}$ (called the conjugation) 
can be constructed as a power series in $\e$ (Lindstedt series)
and for $\e$ small convergence can be proved, exploiting cancellations
and summation methods typical of quantum field theory, \cita{GBG}.
We can call this the {\it maximal} KAM theorem, as it deals
with invariant tori of maximal dimension.

The same methods allow us to study existence of perturbed resonant
quasi--periodic motions in quasi--integrable Hamiltonian systems.
By resonant here we mean that $\Bo_0$ satisfies $1\le s\le d-1$
rational relations, \ie it can be reduced to a vector
$(\Bo,\V0)$, $\Bo\in\RRR^{d-s}$, with rationally--independent components,
via a canonical transformation acting as a linear integer coefficients
map of the angles. In this representation we shall write
$\Bf=(\Ba,\Bb)$ denoting by $\Ba$ the 
``fast variables'' rotating with angular velocity $\Bo$ and by $\Bb$ the 
fixed unperturbed angles (``slow variables''). 

The study of resonant quasi--periodic motions
is mathematically a natural extension of the maximal KAM case and
physically it arises in several stability problems. We mention here
celestial mechanics, where the phenomenon of resonance locking between 
rotation and orbital periods of satellites is a simple example (as in
this case the resonant torus is one--dimensional, \ie it describes a
periodic motion). In general resonant motions arise in presence of
small friction: the most unstable motions are the maximally
quasi--periodic ones (on KAM tori). In presence of friction the
maximally quasi--periodic motions ``collapse'' into resonant motions
with one frequency rationally related to the others, then on a longer
time scale one more frequency gets locked to the others and the motion takes
place on an invariant torus with dimension lower than the maximal by
$2$, and so on. Periodic motions are the least dissipative, and
eventually the motion becomes maximally resonant, \ie periodic. This
is a scenario among others possible, and its study in particular cases
seems to require a good understanding of the properties of the resonant
motions of any dimension.

The first mathematical result is that also in the resonant case,
if $\Bo$ is a $(d-s)$--dimensional Diophantine vector, a conjugation
$\hh_{\e}$ can be constructed by summations of the Lindstedt series;
however the conjugation $\hh_{\e}$ that one is able
to construct is not analytic, but, at best, only $C^\io$ in $\e$:
in general it is defined only on a large measure Cantor set $\EE$
of $\e$'s, $\EE\subset[-\e_0,\e_0]$ for some positive $\e_0$,
so that $C^\io$ has to be meant in the sense of Withney.
It is commonly believed that a conjugation which is analytic in a
domain including the origin does not exist in the resonant case. 
Moreover, contrary to what happens in the maximal case, not all resonant
unperturbed motions with a given rotation vector $\Bo$ appear to survive
under perturbation, but only a discrete number of them.
This is not due to technical limitations of the method,
and it has the physical meaning  that only points $\Bb_0$ which are
equilibria for the ``effective potential'' $(2\p)^{s-d}\int
{\rm d}\Ba \, f(\Ba,\Bb)$ can remain in average at rest 
in presence of the perturbation. 

The following natural (informal) question then arises: ``where do the
unperturbed motions corresponding to initial data $(\Ba,\Bb)$,
$\Bb\neq\Bb_0$, disappear when we switch on the perturbation?''
An intriguing scenario is that the tori that seem to disappear in the 
construction of $\hh_{\e}$ actually ``condense'' into a continuum of 
highly degenerate tori near the ones corresponding to the equilibria
$\Bb_0$. It is therefore interesting to study ``uniqueness'',
regularity and possible degeneracies of the perturbed tori constructed
by the Lindstedt series methods (or by alternative methods,
such as classical Newton's iteration scheme). 

In the present paper we investigate {\it hyperbolic} (see below)
perturbed tori with two-dimensional rotation vectors for 
a class of analytic quasi--integrable Hamiltonians. Informally, 
our main result is that the hyperbolic tori, which survive to the
switching of the perturbation and are described by a function $\hh_{\e}$
that we construct explicitly, are independent on the procedure
used to construct them iteratively (which is not obvious,
due to lack of analyticity). Moreover, if $\h=\sqrt\e$,
{\it the conjugation is Borel summable} in $\h$ for $\e>0$. 
{\it The conjugation constructed here is the unique possible
for our problem within the class of Borel summable functions.} 
Of course this does not exclude existence of other less regular
quasi--periodic motions, not even existence of other quasi--periodic
solutions which admit the same formal power expansion as $\hh_{\e}$.

In order to make our results more precise, we first introduce
the model and summarize the results about existence and properties
of lower--dimensional perturbed tori as we need in the following.
Then we briefly recall the definition and some key properties
of Borel summable functions, and finally we state more
technically our main result. 

\*\sub(1.1) {\it The model.}
Consider a Hamiltonian system 
$$ H=\fra12 {\V A}^2+\fra12{\V B}^2+\h^2 f(\Ba,\Bb) ,
\Eq(1.0)$$ 
with $\II=(\V A,\V B)\in \RRR^r\times\RRR^s$ the action
coordinates and $\Bf=(\Ba,\Bb)\in \TTT^r\times\TTT^s$
the conjugated angle coordinates. Let $\V A_0=\Bo,\V B=\V 0$
be an unperturbed resonance with $|\Bo\cdot\Bn| \ge C_0 |\Bn|^{-\t}$
for $\Bn\in\ZZZ^{r}$, $\Bn\ne\V0$ and for some $C_0,\t>0$, and let
$((\V A_0,\V0),(\Bps+\Bo\,t,\Bb_0))$ be the corresponding
unperturbed resonant motions with initial angles $\Ba=\Bps,\Bb=\Bb_0$.

The following result holds.

\*
\0{\bf Proposition 1.} \cita{GG1} \cita{GG2}
{\it Let $\Bb_0$ be a non--degenerate maximum of the function
$f_{\V0}(\Bb)\=(2\p)^{-r}\int {\rm d} \Ba$ $f(\Ba,\Bb)$, \ie 
$\dpr_\b f_\V0(\Bb_0)=0$ and $\dpr^2_{\Bb} f_\V0(\Bb_0)< 0$,
and let $\Bo\in\RRR^r$ be a Diophantine vector of constants $C_0,\t$,
\ie $|\Bo\cdot\Bn|\ge C_0|\Bn|^{-\t}$ for all $\Bn\in\ZZZ^{r}$,
$\Bn\neq\V0$. Then for $\e>0$, setting $\h=\sqrt{\e}$,
there exists a function $\Bps\to \hh(\Bps,\h)=(\V a(\Bps),\V b(\Bps))$,
vanishing as $\h\to 0$
and with  the following properties.\\
(i) The functions $t\to \Bf(t)= (\Ba(t),\Bb(t))=
(\Bps+\Bo\,t,\Bb_0)+\hh(\Bps+\Bo\,t,\h)$ satisfy
the equation of motion $\ddot\Bf=-\e\dpr_\Bf f(\Bf)$ for any choice of 
$\Bps\in\TTT^r$.\\
(ii) The function $\hh$ is defined for
$(\Bps,\h)\in\TTT^r\times[0,\h_0]$ and can be 
analytically continued to an holomorphic function in the domain
$\DD=\{ \h \in \CCC : \Re\h^{-1}> \h_0^{-1}\}$.\\
(iii) (The analytic continuation of) $\hh$ is $C^\io$ in $\h$ at the
origin along any path contained in $\DD$ and its Taylor coefficients
at the origin satisfy, for suitable positive constants $C$ and $D$,
the bounds $|\fra1{k!}\dpr_\h^k \hh(\Bps,0)|< D C^k k!^\t$,
where $\t$ is the Diophantine exponent of $\Bo$.}\\
\*

\0{\bf Remarks.} \\
(1) The domain $\DD$ is a circle in $\CCC^2$, centered in $\h_0/2$ and 
of radius $\h_0/2$ (hence tangent to the imaginary axis at the origin).
\\
(2) A function $\hh$ was constructed in \cita{GG1} by
a perturbative expansion in power series in $\e=\h^2$ (Lindstedt series) 
and by exploiting multiscale decomposition, cancellations and summations 
in order to control convergence of the series. Eventually  
$\hh$ is expressed as a new series which is not a power series in 
$\e=\h^2$ and which is holomorphic in $\DD$. The procedure leading
to the convergent resummed series from the initial formal power series
in $\e=\h^2$ relies on a number of arbitrary choices
(which will be made explicit in next section) and a priori is not clear
that the result is actually independent on such choices.
Another (in principle) function $\hh$ was constructed in \cita{GG2},
with a method which applies in more general cases (see item (4) below):
we shall see that the two functions in fact coincide.
\\
(3) The function $\hh(\Bps,\h)$ can be regarded as a
function of $\e=\h^2$, as in \cita{GG1} and \cita{GG2}.
As such the bound in item (iii) would be modified into
$|\fra1{k!}\dpr^k_\e \hh(\Bps, 0)| 
< D C^k (2k)!^\t$, and the analyticity domain
in item (ii) would be as described in \cita{GG1}, Fig. 1.
We also note here that the exponent $2\t+1$ in \cita{GG1}, Eq. (5.29),
was not correct (without consequences as the value of the
exponent was just quoted and 
not exploited), as the right one is $3\t$: of course for $\t=1$
(that is the case we consider in this paper)
the two values coincide.
\\
(4) In \cita{GG2} a similar statement was proved for $\Bb_0$ a
non--degenerate equilibrium point of the function $f_{\V0}(\Bb)$,
\ie $\Bb_0$ not necessarily a maximum. If $\Bb_0$ is a maximum 
the corresponding torus is called {\it hyperbolic}, if $\Bb_0$
is a minimum it is called {\it elliptic}. In the elliptic case
for $\e$ real the domain of definition of $\hh$ on
the real line is $\TTT^r\times\EE$, where $\EE\subset[0,\e_0]$ 
is a set with open dense complement in $[0,\e_0]$ but with $0$ as a
density point in the sense of Lebesgue. The reason for stating
Proposition 1 as above is that {\it in the present paper we
shall restrict our analysis to the hyperbolic case}.
\\
(5) If $\t=1$, \ie if $r=2$ and $\Bo$ is quadratically irrational,
than the bound on the coefficients of the power expansion of $h$
in $\h$ at the origin is  $|\hh^{(k)}(\Bps)|< D C^k k!$.
Hence in this case it is natural to ask for Borel summability of $\V h$.
\\
(6) Even if we consider only hyperbolic tori, in the following we
shall use the method introduced in \cita{GG2}, because it is
more general and it is that one should look at if one tried
to extend the analysis to the case of elliptic tori. The main
difference between the forthcoming analysis and that of \cita{GG2}
is the use of a sharp multiscale decomposition,
instead of a smooth one, as it allows further simplifications
in the case of hyperbolic tori. We shall come back to this later.

\*\sub(1.2) {\it Borel transforms.}
Let $F(\h)$ be a function of $\h$ which is analytic in a disk centered
at $(\fra1{2\r_0},0)$ and radius $\fra1{2\r_0}$ (\ie centered on the
positive real axis and tangent, at the origin, to the imaginary axis),
and which vanishes as $\h\to0$ as $\h^q$ for some $q>1$. Then one
can consider the inverse Laplace transform of the function $z\to
F(z^{-1})$, defined for $p$ real and positive and $\r>\r_0$ by
$$ \LL^{-1}F(p)=\ig_{\r-i\io}^{\r+i\io} e^{z\,p} F\big(\fra1z \big) \,
\fra{{\rm d}z}{2\p i}.
\Eq(2.1) $$
If $F$ admits a Taylor series at the origin in the form
$F(\h)\sim\sum_{k=2}^\io F_k \h^k$ then the Taylor series
for $\LL^{-1}F$ at the origin is
$$ \LL^{-1}F(p)\sim\sum_{k=2}^\io F_k \fra{p^{k-1}}{(k-1)!} .
\Eq(2.2) $$
If the series in \equ(2.2) is convergent then the sum of the series
coincides with $\LL^{-1}F(p)$ for $p>0$ real. Of course the series
expansion \equ(2.2) provides an expression suitable for studying
analytic continuation of $\LL^{-1}F(p)$ outside the real positive
axis.  Note that \equ(2.2) makes sense even in the case the series
starts from $k=1$. Then given any formal power series $F(\h)\sim
\sum_{k=1}^\io F_k\h^k$ we shall define $F_B(p)= \sum_{k=1}^\io F_k
\fra{p^{k-1}}{(k-1)!}$ as the {\it Borel transform} of $F(\h)$,
whenever the sum defining it is convergent.  It is remarkable that in
some cases the map $F\to F_B$ is invertible.  If this is the case one
says that the Taylor series of $F$ is {\it Borel summable} and we also
say that the function $F$ is {\it Borel summable}: this is made
precise as follows.

\*
\0{\bf Definition 1.} {\it Let a function $\h\to F(\h)$ be such that
\\
(i) it is analytic in a disk centered at $(\fra1{2\r_0},0)$ and radius
$\fra1{2\r_0}$, and admits an asymptotic Taylor series at the origin
where it vanishes,
\\
(ii) its Taylor series at the origin admits a
Borel transform $F_B(p)$ which is analytic for $p$ in a neighborhood
of the positive real axis, and on the positive real axis grows
at most exponentially as $p\to+\io$,
\\
(iii) it can be expressed, for $\h>0$ small enough, as
$$ F(\h)=\ig_0^{+\io} e^{-p/\h} F_B(p) \, {\rm d} p .
\Eq(2.3)$$
Then we call $F$ {\rm Borel summable} (at the origin).}
\*

\0{\bf Remarks.}\\
(1) If $F$ is Borel summable, then one says that 
$F$ is equal to the Borel sum of its own Taylor series.
\\
(2) For instance one checks that a function $F$ holomorphic
at the origin and vanishing at the origin is Borel summable;
its Borel transform is entire.
\\
(3) The function $F(\h)=\sum_{k=1}^\io 2^{-k}\fra{\h}{1+k\h}$ is not
analytic at the origin but it is Borel summable.
\\
(4) If $F,G$ are Borel summable then also $FG$ is Borel summable 
and $(FG)_B(p)=\ig_0^p F_B(p')G_B(p-p')dp'\= (F_{B}*G_{B})(p)$
on the common analyticity domain of $F_B$ and $G_B$,
with the integral which can be computed along any path from $0$ to $p$ 
in the common analyticity domain. Of course by definition 
$|F_B*G_B|\le |F_B|*|G_B|$, where in the r.h.s. the convolution is along
any path from $0$ to $p$ in the common analyticity domain of $F_B$
and $G_B$ (note however that now the convolution $|F_B|*|G_B|$
depends on the path).
\\
(5) The Borel transform of $\h^k$ is $p^{k-1}/(k-1)!$.
The Borel transform of $\h/(1-\a\h)$ is $e^{\a p}$. If $F_B(p)=
e^{\a p}p^{k_1}/k_1!$ and $G_B(p)=e^{\a p}p^{k_2}/k_2!$, then
$F_B*G_B(p)=e^{\a p}p^{k_1+k_2+1}/(k_1+k_2+1)!$
\\
(6) More generally if $F_i$, $i=1,2$, have Borel transforms $F_{i\,B}$
bounded in a sector around the real axis and centered at the origin by
$|F_{i\,B}(p)|\le C e^{\a_i|p|+\b_i|{\rm Im}\,p|}
|p|^{k_i}/k_i!$, with $\a_i$ and $\b_i$ real,
then $|F_{1\,B}*F_{2\,B}(p)|\le C^2 e^{\a |p| + \b |{\rm Im}\,p|}
|p|^{k_1+k_2+1}/(k_1+k_2+1)!$ where $\a=\max \a_i$
and $\b=\max \b_i$. We shall make use of this bound repeatedly below.

\*\sub(1.3) {\it Main results}.
We are now ready to state more precisely our main results.

\*
\0{\bf Proposition 2.} {\it Let us consider Hamiltonian \equ(1.0) 
with $r=2$ and $f(\Ba,\Bb)$ an analytic function of its arguments.
Let $\Bb_0\in\TTT^s$ satisfy the assumptions of Proposition 1
and $\Bo\in\RRR^2$ be a Diophantine vector with constants $C_0>0$
and $\t=1$, \ie $|\Bo\cdot\Bn| \ge C_0|\Bn|^{-1}$
for all $\Bn\in\ZZZ^{r}$, $\Bn\ne\V0$. Then there exists
a unique Borel summable function $\hh(\Bps,\h)$ satisfying properties 
(i)--(iii) of Proposition 1.}
\*

Note that, once existence of a Borel summable function $\hh(\Bps,\h)$
satisfying properties (i)--(iii) of Proposition 1 is obtained, 
the uniqueness in the class of Borel summable functions is obvious,
by the very definition of Borel summability. In fact all such
functions have a Borel transform $\hh_B(\Bps,p)$ that is $p$--analytic
in an open domain enclosing $\RRR^+$ (hence also in a neighborood
of the origin), where they coincide (because they all have the same
expansion at the origin), then they all coincide everywhere.

The proof of Proposition 2 will proceed by showing Borel summability 
of (one of) the function(s) $\hh(\Bps,\h)$ constructed in \cita{GG2}.
In particular a corollary of the proof is that $\hh(\Bps,\h)$
constructed in \cita{GG2} is independent of the arbitrary 
choices mentioned in Remark (2) after Proposition 1.

Our proof of Borel summability of $\hh$ does not use Nevanlinna's
theorem, \cita{Ne} \cita{So}. In fact we failed in checking that $\hh$
satisfies the hypothesis of the theorem. Our strategy goes as follows.
We introduce a sequence of approximants $\hh^{(N)}$ to $\hh$, 
that is naturally induced by the multiscale construction of
\cita{GG2}. We explicitly check that $\hh^{(1)}$
is Borel summable and that its Borel transform is entire.  
Then we show inductively that the analyticity domain of $\hh^{(N)}_B$
is a neighborood $\BBB$ of $\RRR^+$ (not shrinking to $0$ as 
$N\to\io$) and that $\hh^{(N)}_B$ grows very fast at infinity
in $\BBB$ (in general faster than exponential). However the results
of Proposition 1 imply that the growth of $\hh^{(N)}_B$ 
{\it on the positive real line} is uniformly bounded by an exponential.
Then Borel summability of $\hh$ follows by performing the limit
$N\to\io$ and using uniform bounds that we shall derive
on the approximants and on their Borel transforms.

In the next section we will recall the structure and the properties of the 
resummed series obtained in \cita{GG2}, defining the function 
$\hh(\Bps,\h)$ of Proposition 1. In Section \sec(4) we define the sequence 
$\hh^{(N)}$ of approximants and we show that the inverse Laplace transform
of $\hh^{(N)}$ is uniformly bounded by an exponential on the positive real 
line. In Section \sec(5) we prove Borel summability of $\hh(\Bps,\h)$ in 
the easier case in which the perturbation $f(\Ba,\Bb)$ in \equ(1.0)
is a trigonometric polynomial in $\Ba$. In Appendix \sec(A1) we discuss
how to extend the method to cover the general analytic case.
Finally, in Appendix \sec(A2) we show that the same result applies
to the function $\hh$ constructed in \cita{GG1}: this allows us
to identify the functions constructed with the two methods
of \cita{GG1} and \cita{GG2}, since they are both Borel summable
and admit the same formal expansion at the origin.

\*\*
\section(3,Lindstedt series)

\0Denote by $\V a(\Bps),\V b(\Bps)$ the $\Ba,\Bb$ components of $\hh$,
respectively. In \cita{GG2} an algorithm is described to
construct order by order in $\h^2$ the solution to the homologic equation
$$\cases{(\Bo\cdot\dpr_\Bps)^2\V a(\Bps)=\h^2 \dpr_\Ba 
f(\Bps+\V a(\Bps),\Bb_0+\V b(\Bps)) , \cr
(\Bo\cdot\dpr_\Bps)^2\V b(\Bps)=\h^2 \dpr_\Bb 
f(\Bps+\V a(\Bps),\Bb_0+\V b(\Bps)) . \cr}
\Eq(2.0) $$
The resulting series, called the ``Lindstedt series'',
is widely believed to be divergent. A summation
procedure has been found which collects its terms into families until
a convergent series is obtained. The resummed series (no longer a
power series) can be described in terms of suitably decorated tree
graphs, \ie $\hh$ can be expressed as a sum of values of tree graphs:
$$ h_{\g,\Bn} (\h)= \sum_{\th \in \Th_{\Bn,\g}} \Val(\th) ,
\Eq(3.3)$$
where $\hh_\Bn$ is the $\Bn$--th coefficient in the Fourier series for
$\hh$, and $\g=\{1,\ldots,2+s\}$ labels the component of the vector
$\hh_\Bn$ (recall that $2+s$ is the number of degrees of freedom of
our Hamiltonian, $2$ being the number of ``fast variables'' $\Ba$
and $s$ being the number of ``slow variables'' $\Bb$).
$\Th_{\Bn,\g}$ is the set of decorated trees
contributing to $h_{\g,\Bn} (\h)$ and, given $\th \in \Th_{\Bn,\g}$,
$\Val(\th)$ is its value, both still to be defined.

We now describe the rules to construct the tree graphs and to 
compute their value. We shall need the explicit structure 
in the proof of Borel summability in next sections, and this is why we are
reviewing it here. Given the rules below one can formally check that
the sum \equ(3.3) is a solution to the Hamilton equations,
see \cita{GG2}. A few differences (in fact simplifications)
arise here with respect to \cita{GG2}, and we provide some
details with the aim of making the discussion self-consistent.
Essentially, the changes consist of: (i) shifting the order of factors
in products appearing in the definition of the values $\Val(\th)$
to an order that makes it easier to organize the recursive evaluation of
several Borel transforms; see remarks following \equ(3.8), and
(ii) using a sharp multiscale decomposition; see item (f) below.

Consider a tree graph (or simply tree) $\th$ with $k$ nodes
$v_1,\ldots,v_k$ and one root $\V r$, which is not considered
a node; the tree lines are oriented towards the root (see Fig.1).

\*
\eqfig{200pt}{141.6pt}{
\ins{-11pt}{72pt}{$\V r$}
\ins{14pt}{71pt}{$\Bn\kern-2pt=\kern-2pt\Bn_{\ell_{0}}$}
\ins{24pt}{91pt}{$\ell_{0}$}
\ins{50pt}{71pt}{$v_{0}$}
\ins{46pt}{91pt}{$\Bn_{v_{0}}$}
\ins{86pt}{105pt}{$\h$}
\ins{127pt}{100.pt}{$v_{1}$}
\ins{191.7pt}{133.3pt}{$v_{5}$}
\ins{156.7pt}{138pt}{$\h'$}
\ins{121.pt}{125.pt}{$\Bn_{v_{1}}$}
\ins{92.pt}{43.6pt}{$v_{2}$}
\ins{158.pt}{83.pt}{$v_{3}$}
\ins{191.7pt}{100.pt}{$v_{6}$}
\ins{191.7pt}{71.pt}{$v_{7}$}
\ins{191.7pt}{-6.0pt}{$v_{11}$}
\ins{191.7pt}{18.2pt}{$v_{10}$}
\ins{166.6pt}{42.pt}{$v_{4}$}
\ins{191.7pt}{54.2pt}{$v_{8}$}
\ins{191.7pt}{37.5pt}{$v_{9}$}
\ins{32pt}{139pt}{$\h$}
\ins{53pt}{130pt}{$\g$}
\ins{12pt}{134pt}{$\g'$}
\ins{63pt}{115pt}{$v$}
\ins{0pt}{117pt}{${v}'$}
}{fig1}{}
\*\*
\line{\vtop{\line{\hskip1.5truecm\vbox{\advance\hsize by -4.1 truecm
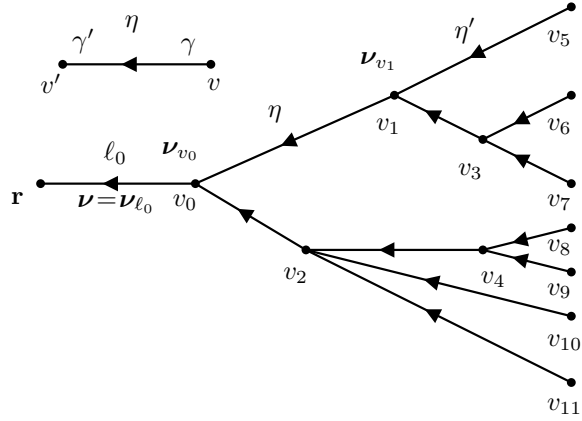
\0{{\css Figure 1.} {\nota A tree $\th$ with $12$ nodes;
one has $p_{v_{0}}=2,p_{v_{1}}=2,p_{v_{2}}=3,
p_{v_{3}}=2, p_{v_{4}}=2$.  The length of the lines should
be the same but it is drawn of arbitrary size.
The separated line illustrates the way to think of the label
$\h=(\g',\g)$.\vfil}} } \hfill} }}
\*
\*

\kern-4mm

The line entering the root is called the root line.
We denote by $V(\th)$ and $\L(\th)$ the set of nodes
and the set of lines in $\th$, respectively.

\*

\0(a) On each node $v$ a label $\Bn_v\in\ZZZ^2$,
called the {\it mode label}, is appended.
\\
(b) To each line $\ell$ a pair of labels $\h=(\g',\g)$
is attached. $\g'$ and $\g$ 
are called the left or right {\it component labels},
respectively: $\g'\in (1,\ldots,2+s)$ is associated with the left
endpoint of $\ell$ and $\g\in (1,\ldots,2+s)$ with the right endpoint
(in the orientation toward the root, see Fig.1). The label $\g'$
associated with the root line will be denoted by $\g(\th)$. 
\\
(c) Each node $v$ will have $p_{v} \ge 0$ {\it entering lines}
$\ell_{1},\ldots,\ell_{p_{v}}$. Hence with the node
$v$ we can associate the left component labels
$\g_{v 1}',\ldots,\g_{v p_{v}}'$ of the entering lines,
and an extra label $\g_{v 0}=\g_{\ell}$ attached
to the right endpoint of the line exiting from $v$.
Thus a tensor $\dpr_{\g_{v 0}\g_{v 1}\cdots\g_{v p_{v}}}
f_{\Bn_{v}}(\Bb_0)$ can be associated with each node $v$,
with $\dpr_\g$ denoting the derivative with respect to $\b_\g$
if $\g>2$ and multiplication by $i\n_\g$ if $\g\le 2$.
\\
(d) A {\it momentum} $\Bn_{\ell}$ is associated with each line
$\ell=v'v$ oriented from $v$ to $v'$: this
is a vector in $\ZZZ^r$ defined as $\Bn_{\ell}=
\sum_{w\le v} \Bn_{w}$. The {\it root momentum},
that is the momentum through the root line, will be denoted by $\Bn(\th)$.
\\
(e) A {\it number label} $k_\ell\in\{1,\ldots,|\L(\th)|\}$ is associated
with each line $\ell$, with $\cup_{\ell\in\L(\th)}\{k_\ell\}= \{1,\ldots,$
$|\L(\th)|\}$. The number label is used for combinatorial purposes: 
two trees differing only because of the number labels are still
considered distinct.
\\
(f) Each line $\ell$ also carries a {\it scale label}
$n_{\ell}=-1,0,1,2,\ldots$: this is a number which determines
the size of the {\it small divisor} $\Bo\cdot\Bn_{\ell}$,
in terms of an exponentially decreasing sequence $\{\g_p\}_{p=0}^\io$
of positive numbers that we shall introduce in a moment.
If $\Bn_\ell=\V 0$ then $n_\ell=-1$.
If $|\Bo\cdot\Bn_{\ell}|\ge \g_0$ then $n_\ell=0$, and we say that
the line $\ell$ (or else $\Bo\cdot\Bn_{\ell}$) is on scale $0$.
If $\g_p\le |\Bo\cdot\Bn|< \g_{p-1}$ for some $p$ then $n_\ell=p$,
and we say that the line $\ell$ (or else $\Bo\cdot\Bn_{\ell}$)
is on scale $n$. 
The sequence $\{\g_p\}_{p=0}^\io$ is such that
$\g_p \in C_0[2^{-p-2},2^{-p-1})$ for all $p\ge 0$
and, furthermore, $\Bo\cdot\Bn$ not
only stays bounded below by $C_0|\Bn|^{-1}$ (because of the Diophantine 
condition) but it stays also ``far''
from the values $\g_{p}$ for $\Bn$ not too large, \ie for
$|\Bn|$ at most of order $2^{p}$; cf. \cita{GG} for a proof of the existence 
of the sequence (without further assumptions on $\Bo$). Precisely,
$$ \eqalign{
(1)\kern1.truecm&
|\Bo\cdot\Bn|\ge C_0|\Bn|^{-1},
\kern3.5cm\V0\ne\Bn\in \ZZZ^r , \cr
(2)\kern1.truecm&
\min_{0\le p\le n}\big| |\Bo\cdot\Bn|-\g_{p}\big| > C_0 2^{-n}
\qquad {\rm if}\ n\ge0,\ 0<|\Bn|\le 2^{(n-3)} . \cr}
\Eq(3.1) $$
Note that the definition of scale of a line depends on the arbitrary choice
of the sequence $\{\g_p\}$: we could as well used a sequence scaling as 
$\g_p\sim \g^{-p}$, with $\g$ any number $>1$ instead of $\g=2$; or we 
could have used a smooth cutoff function (as in \cita{GG2}) replacing the 
sharp cutoff function $\openone(\g_p\le |\Bo\cdot\Bn|< \g_{p-1})$
implied in the definition above.
\\
(g) The scale labels allow us to define hierarchically
ordered clusters. A cluster $T$ of scale $n$ is a maximal connected set
of lines $\ell$ on scale $n_{\ell}$, with $n_{\ell}\le n$,
containing at least one line on scale $n$. The lines which
are connected to a line of $T$ but do not belong to $T$ are called
the {\it external lines} of $T$: according to their orientations,
one of them will be called the {\it exiting line} of $T$,
while all the others will be the {\it entering lines} of $T$.
All the external lines $\ell$ are on scales
$n_{\ell}$ with $n_{\ell}>n$. The set of lines of $T$,
called the {\it internal lines} of $T$, will be denoted by $\L(T)$
and the set of nodes of $T$ will be denoted by $V(T)$.
\\
(h) Not all arrangements of the labels are permitted. The ``allowed
trees'' will have no nodes with $\V0$ momentum and with only
one entering line and the exiting line also carrying $\V0$ momentum.
We also discard trees which contain clusters with only one entering line
and one exiting line with equal momentum and with no line with $\V0$
momentum on the path joining the entering and exiting lines 
(``self--energy'' clusters or ``resonances'').\\

\*
\0{\bf Remark.}
One can verify that chains of self--energy clusters can actually appear 
in the initial formal Lindstedt series. 
One of the main points in \cita{GG1} and \cita{GG2} is to
show that if one modifies the series by descarding all chains of
self--energy clusters, then the resulting series is convergent
(a form of Bryuno's lemma that appears in KAM theory). 
In both \cita{GG1} and \cita{GG2} it is shown that, in order to deal
with chains of self--energy diagrams one can iteratively resum them
into the propagators (\ie the factors associated with the tree lines
in the value of a tree, see below for a definition), that will then
turn out to be different from those appearing in the naive formal
Lindstedt series (which are simply $(\Bo\cdot\Bn)^{-2}$).
Such resummation is the analogue of Dyson's equations in
quantum field theory and the iteratively modifed propagator has been,
therefore, called the {\it dressed propagator}. Here there is
further freedom in the choice of the self--energy clusters.
The idea is that the self--energy clusters must 
include the ``diverging contributions'' affecting the initial 
formal power series. But if we change the definition of
self--energy clusters by adding to the class a new class of
non-diverging clusters, the construction can be shown to go through
as well. We find convenient the specific choice above but this is
of course not necessary. This is the second arbitrary choice
we do in the iterative construction of the resummed series.
It can fuel doubts about the uniqueness of the result which
can only be dismissed by further arguments (like the Borel
summability that we are proving).
\*

The set of all allowed trees with labels $\g(\th)=\g$ and $\Bn(\th)=\Bn$
is denoted by $\Th_{\Bn\g}$ (this is the set appearing in \equ(3.3)). 
The labels described above are used to define the {\it value} 
$\Val(\th)$ of a (decorated) tree $\th\in\Th_{\Bn\g}$: 
this is a number obtained by multiplying the following factors:

\0(1) a factor $F_{v}=\dpr_{\g_{v 0}\g_{v 1}'\cdots
\g_{v p_{v}}'}f_{\Bn_{v}}(\Bb_0)$,
called the {\it node factor}, per each node $v$;
\\
(2) a factor $g^{[n_{\ell}]}_{\ell}=g^{[n_{\ell}]}_{\g'_{\ell},\g_{\ell}}
(\Bo\cdot\Bn_{\ell};\h)$, called the {\it propagator},
per each line of scale $n_{\ell}$, momentum $\Bn_{\ell}$
and component labels $\g'_{\ell},\g_{\ell}$, see items (I)--(V)
below for a definition.

The value is then defined as
$$ \Val(\th)= \fra1{k!} \big(\prod_{v \in V(\th)} F_{v} 
\big) \big(\prod_{\ell \in \L(\th)}
g^{[n_{\ell}]}_{\ell} \big) ,
\Eq(3.2)$$
where it should be noted that all labels $\g$ (of the tensors
$F_{v}$ and of the matrices $g^{[n_{\ell}]}_{\ell}$)
appear repeated twice because they appear in the propagators as well
as in the tensors associated with the nodes, with the exception of the
label $\g$ associated with the left endpoint of the line ending in the
root (as the root is not a node and therefore there is no tensor
associated with it).

Adopting the convention of summation over repeated component labels
$\Val(\th)$ depends on the root label $\g(\th)=
\g=1,\ldots,2+s$ so that it defines a vector in $\CCC^{2+s}$.

The recursive definition of the propagators is such that the series
in \equ(3.3) is convergent and gives the $\Bn$-th Fourier component
of the function $\V h(\Bps,\h)$ in Sect. \sec(1).
The definition of propagators we adopt here is the same introduced in
\cita{GG2}: the definition in \cita{GG1} is slightly different
(see Appendix \sec(A2)), but it has the drawback that it is specific
for hyperbolic resonances, while the definition in \cita{GG2} can be
(expected to be) extended also to the theory of elliptic resonances
and, therefore, might turn out to be useful in view of possible
extensions of the main results of this work to elliptic resonances.

\*

\0(I) For $n=-1$ the propagator of the line $\ell$ 
is defined as the block matrix
$$ g^{[-1]}_{\ell} \defi
\pmatrix{0&0\cr0&(- \dpr^2_{\Bb} f_\V0(\Bb_0))^{-1}\cr} .
\Eq(3.4)$$
(II) For $n=0$, if the line $\ell$ carries a momentum $\Bn$ and if
$x\defi \Bo\cdot\Bn$, the propagator is the matrix
$$g^{[0]}_{\ell} = g^{[0]}(x;\h) =
\fra{\h^2}{x^2+\h^2 M_0} ,
\Eq(3.5)$$
with $M_0\defi \pmatrix{0&0\cr0 & -\dpr^2_{\Bb} f_\V0(\Bb_0)\cr}$.
{\it By the assumptions of Proposition 2 one has $M_0\ge 0$}.
\\
(III) For $n>0$ the propagator is the matrix
$$ g^{[n]}_{\ell} = g^{[n]}(x;\h) =
\fra{\h^2}{x^2+\MM^{[\le n]}(x;\h)} .
\Eq(3.6)$$
with $\MM^{[\le n]}(x;\h) = \MM^{[0]}(x;\h) + \MM^{[1]}(x;\h) +
\ldots + \MM^{[n]}(x;\h)$, where $\MM^{[0]}(x;\h) = \h^2 M_0$,
whereas $\MM^{[j]}(x;\h)$, $j\ge 1$, are
matrices, called {\it self--energy matrices}, whose expansion in $\h$
starts at order $\h^4$ and are defined as described in the next two items.
\\
(IV) Let $T$ be a self--energy cluster on scale $n$ (see item (h) above)
and let us define the matrix\footnote{${}^1$}{\nota 
This is a matrix because the self--energy cluster inherits the
labels $\g,\g'$ attached to the left of the entering line and
to the right of the exiting line.\vfil} $\VV_{T}(\Bo\cdot\Bn;\h)$ as
$$ \VV_{T}(\Bo\cdot\Bn;\h) = -{\h^2\over |\L(T)|!}
\Big(\prod_{v \in V(T)} F_{v} \Big)
\Big(\prod_{\ell \in \L(T)} g^{[n_{\ell}]}_{\ell} \Big) ,
\Eq(3.7) $$
where, necessarily, $n_{\ell}\le n$ for all $\ell\in \L(T)$.
The matrix \equ(3.7)
will be called the {\it self--energy value} of $T$. The set of the
self--energy clusters with value proportional to $\h^{2k}$,
hence with $k-1$ internal lines with $n_\ell\ge 0$, 
and with maximum scale label $n$ will be denoted $\SS^{\RR}_{k,n}$.
\\
(V) The {\it self--energy matrices} $\MM^{[n]}(x;\h)$, $n\ge 1$,
are defined recursively for $|x| \le \g_{n-1}$ (i.e. for
$x$ on scale $\ge n$) as
$$ \MM^{[n]}(x;\h) = \sum_{k=2}^{\io}
\sum_{T \in \SS^{\RR}_{k,n-1} } \VV_{T}(x;\h) ,
\Eq(3.8)$$
where the self--energy values are evaluated by means of the
propagators on scales $p$, with $p=-1,0,1,\ldots,n-1$.

\*
\0{\bf Remarks.}\\
(1) With respect to \cita{GG2} the second argument
of the propagators (and of the self--energy values and matrices)
has been denoted $\h$ instead of $\e$; we recall that the variable
$\h^2$ appearing here is the same as the variable $\e$ appearing
in \cita{GG1} and \cita{GG2}. We make this choice because it is
natural to study Borel summability in $\h$ and not in $\e=\h^2$. 
\\
(2) The association of the factors $\h^{2}$ with the lines themselves
rather than with the nodes (as in \cita{GG1} and \cita{GG2}) will be
more convenient when considering the Borel transforms of the involved
quantities.
\\
(3) The multiscale decomposition used in \cita{GG2}
may look quite different from the one we are using here,
but this is not really so. First, even though the decomposition
in \cita{GG2} was based on the propagator divisors
$\D^{[n]}(x;\e)= \min_j|x^{2}-\ul{\l}^{[n]}_{j}(\e)|$, where
the self--energies $\ul{\l}^{[n]}_{j}(\e)$ were defined
recursively in terms of the self--energy matrices,
in the case of hyperbolic tori one has identically
$\D^{[n]}(x;\e)=x^{2}$: indeed all self--energies which are non-zero
are strictly negative. Then the only real difference is that
here we are using a sharp decomposition instead of a smooth one, 
but the latter is not a relevant difference. In fact the choice of
the sequence $\{\g_{p}\}_{p=0}^{\io}$ implies that the lines
appearing in the groups of graphs that will be collected together to
exhibit the necessary cancellations have currents $\Bn$ such that
$\Bn\cdot\Bo$ stays relatively far from the extremes of the intervals
$[\g_{p+1},\g_p]$ that define the scale labels,
and this allows us to use a sharp multiscale decompositions instead of
the smooth one used in \cita{GG2}. In other words this change with respect
to \cita{GG2} is done only to avoid introducing partitions of unity by
smooth functions and the related discussions.

\*

Therefore the expression \equ(3.3) makes sense and  in fact the function
$\hh$ mentioned in Proposition 1 is exactly the Fourier sum of the r.h.s.
of \equ(3.3). In particular in \cita{GG2} it was
proved that the Fourier sum of the r.h.s. of \equ(3.3) satisfies
the properties (i)--(iii) in Proposition 1. 

\*\*\section(4,Integral representation of the resummed Lindstedt series)

\0Given the definitions of Sect. \sec(3) consider the function
${\V h}^{(N)}$ defined in the same way as $\V h$ but restricting the sum
in \equ(3.3) to the trees containing only lines of scale $n\le N$.
 
The functions $\V h^{(N)}$ have the ``same'' convergence and
analyticity properties of the functions $\V h$ and the same bounds on
the Taylor coefficients at the origin. Moreover $\V h^{(N)}
\tende{N\to\io} \V h$: this is a consequence of the intermediate steps
in the proof of the above proposition in \cita{GG2},
as the strategy of the proof is to define $\V h^{(N)}$ making sure
that the convergence and analyticity properties are uniform in $N$. In
fact $\V h^{(N)}$ is even analytic in $\h$ near the origin for
$|\h|\le\h_N$ (but $\h_N\tende{N\to\io}0$).

Therefore the functions $\V h^{(N)}$ are trivially Borel summable,
and have an entire Borel transform, but the growth at $p\to+\io$ of
their Borel transforms is $N$--dependent while, to show Borel
summability of $\V h$, uniform estimates are needed. This section is
devoted to a first attempt at such bounds which uses minimally the
informations on the resummed series that can be gathered from
\cita{GG2}, \ie the convergence properties just mentioned.

The Borel transform of the functions $\V h^{(N)}(\Bps,\h)$ is
an entire function that can be written for $p$ real and positive as
$$ (\V h^{(N)})_B(\Bps,p)=\LL^{-1}h(\Bps,p)=
\ig_{\h^{-1}_N-i\io}^{\h_N^{-1}+i\io}
{\rm e}^{pz} \, \V h^{(N)}\big(\Bps,\fra1z\big) \fra{{\rm d}z}{2\p i} ,
\qquad\qquad p\in\RRR^+ ,
\Eq(4.1) $$
where $\h_N$ is the convergence radius of $\hh^{(N)}(\Bps,\h)$.  
The key remark is that we also know that by property (ii) 
in Proposition 1 (actually by the same property for $\hh^{(N)}$
that follows from the construction in \cita{GG2})
the function $\V h^{(N)}(\Bps,\fra1z)$ is analytic for $|z|>2\h_0^{-1}$, 
so that the integral in \equ(4.1) can be shifted to a
contour on the vertical line with abscissa $\lis\r>2\h_0^{-1}$, \ie
with $N$-independent abscissa. Therefore
$$ (\V h^{(N)})_B(\Bps,p)=
\ig_{\bar\r-i\io}^{\bar\r+i\io}
{\rm e}^{pz} \, \V h^{(N)} \big(\Bps,\fra1z \big)
\fra{{\rm d} z}{2\p i} ,\qquad\qquad p\in\RRR^{+} ,
\Eq(4.2) $$
and, for all $N$, the function $\V h^{(N)}(\Bps,\fra1z)$ is uniformly
bounded by $O(\fra1{|z|^2})$ on the integration contour, because,
by property (iii) in Proposition 1, $\hh$ is twice differentiable
at the origin along any path contained in $\DD$
(in particular along the circular path $\Re\h^{-1}=\r$).
Hence the latter boundedness property of $\V h^{(N)}(\Bps,\fra1z)$
and \equ(4.2) imply the bound, for $p>0$ and for a suitable constant $C$,
$$ |(\V h^{(N)})_B(\Bps,p)|\le \, C\,e^{\bar\r p} ,
\qquad\qquad \forall p\in\RRR^{+} ,
\Eq(4.3) $$
and for all $\bar\r>2\h_0^{-1}$. The existence of the
limit $\lim_{N\to\io} \V h^{(N)}(\Bps,\fra1z)=\V h(\Bps,\fra1z)$
for $\fra1{|z|}$ small (by \cita{GG2}) implies existence of the limit
$\V F(\Bps,p)$ as $N\to\io$ of $(\V h^{(N)})_B(\Bps,p)$ for
$p\in\RRR^+$ and $\V F(\Bps,p)$ satisfies the bound \equ(4.3) on $\RRR^+$.

Hence the functions $\V h(\Bps,\h)$ can be expressed,
for $0\le \h<\h_0$, as
$$ \V h(\Bps,\h)=\ig_0^\io e^{-p/\h}\, \big(\lim_{N\to\io}
(\V h^{(N)})_B(\Bps,p)\big)\, {\rm d} p
=\ig_0^\io e^{-p/\h} \, \V F(\Bps,p)\,{\rm d}p ,
\Eq(4.4) $$
which provides us with an integral representation of the resummed
series and shows that the resummation \equ(3.3) generates a Borel sum
of the formal Lindstedt series {\it provided} $\V F(\Bps,p)$ can be
shown to be analytic in a neighborhood of the positive axis, as required
by the very definition of Borel summability, see property (ii) in 
Definition 1 of Section \sec(1.2).

\*\*
\section(5, Borel summability)

\0By the remark at the end of Sect. \sec(4) Borel summability of $\V h$
will be established once the natural candidate for its Borel
transform, namely the function $\V F$ in \equ(4.4), is shown to be
analytic in a region containing the positive real axis. This will be
done here assuming, for simplicity, that the perturbation $f$ is a
trigonometric polynomial. The general case of an analytic $f$ is
slightly more involved and will be treated in Appendix \sec(A1).

Assume, inductively, that $\forall x=\Bo\cdot\Bn$ the functions
$(\MM^{[\le k]})_B(x;p)$ are entire functions of $p$ and 
$$ \big\| \big(\MM^{[\le k]} \big)_B(x;p)\big\|\le \G|p|
e^{ d_k\, |p|}, \qquad {\rm for\
all}\ x \ {\rm on \ scale}\ k', \ {\rm with} \ k' \ge k\ge 0 ,
\Eq(5.1) $$
for all $p$ complex. The matrix norm in \equ(5.1) is $||\MM||=
\max_j \sum_i|\MM_{ij}|$.
Note that $(\MM^{[0]})_{B}(p)=p M_{0}$ (so that the inductive assumption
in \equ(5.1) is valid at the first step $k=0$ with $d_0=0$ if $\G\ge
||M_0||$) 
and $(g^{[0]})_B(x;p)$ is entire and it is given by
$$(g^{[0]})_B(x;p)=\fra1{x^2} \fra{\sin p\sqrt{M_0x^{-2}}}{
\sqrt{M_0x^{-2}}}\quad\tto\quad 
\|(g^{[0]})_B(x;p)\|\le \fra{|p|}{x^{2}}
e^{|p|{c_0}} ,\quad\forall |x|\ge \g_0 ,
\Eq(5.2) $$
with $c_0=\sqrt{||M_0||}/\g_0$.

Supposing the inductive assumption to be valid for $k=1,\ldots,N-1$
we remark that this implies a bound on $(g^{[k]})_{B}(x;p)$ for
$k=1,\ldots,N-1$ via the expansion
$$ (g^{[k]})_B(x;p)= \Big(\fra{\h^2}{x^2} \sum_{m=0}^\io \Big(\fra{
-\MM^{[\le k]}(x;\cdot)}{x^2}\Big)^m\Big)_B .
\Eq(5.3) $$
Taking the Borel transform and performing all convolutions along a straight
line from $0$ to $p$, for $x$ on scale $1\le k\le N-1$, we get
$$\eqalign{
\|(g^{[k]})_B(x;p)\|&\le \fra1{x^2}
\sum_{m=0}^{\io} |p|*\fra{(\G|p|e^{d_k|p| })^{*m}}{x^{2m}}
\le \fra{|p|}{x^2} \sum_{m=0}^{\io} e^{d_k|p|}
\fra{\G^{m}|p|^{2m}}{(2m)!}\fra1{x^{2m}}\le\cr
& \le  \fra{|p|}{x^2}
e^{d_k|p|+\G^{1/2}\g_k^{-1}|p|}\=\fra{|p|}{x^2} e^{c_k|p|}\cr} ,
\Eq(5.4) $$
with $f^{*m}\defi f*\ldots*f$ $m$ times and $c_k=d_k+\G^{1/2}\g_k^{-1}$.

Then $(\MM^{[\le N]} - \MM^{[0]})_B(x;p)$
is estimated via \equ(3.8), that is
$$\eqalign{& \| (\MM^{[\le N]} - \MM^{[0]})_B(x;p) \|
\le\cr
&\le \sum_{k=2}^\io\
\sum_{T\in \cup_{j=0}^{N-1}
\SS^{\RR}_{k,j}}
\fra1{|\L(T)|!}|p|*\Big(\mathop{{\prod}^{*}}_{\ell\in\L(T)\atop n_\ell\ge 0} 
\fra1{x_{\ell}^2}|p|
e^{c_{n_{\ell}} |p|}\Big)\Big(\prod_{\ell\in\L(T)\atop n_\ell=-1} 
\|g_\ell^{[-1]}\|\Big)
\Big(\prod_{v\in V(T)} \| F_{v}\| \Big) , \cr}
\Eq(5.5) $$
where $x_{\ell}=\Bo\cdot\Bn_{\ell}$, the $\prod^*$ is a convolution product
and 
$$ \|F_v\|=\max_{\g'_{v1},\g'_{v2},\ldots,\g'_{v p_v}}\sum_{\g_{v0}}
|(F_v)_{\g_{v0},\g'_{v1}\g'_{v2}\ldots\g'_{v p_v}}| .
\Eq(5.5a) $$
Hence bounding $|p|*\Big(\prod^*|p|e^{c_{n_{\ell}} |p|}\Big)$
by $e^{c_{N-1} |p|} |p|^{2k-1}/(2k-1)!$ and using the estimates in
\cita{GG2} to control the sum over the
self--energy clusters, the bound becomes

$$ \| (\MM^{[\le N]} )_B(x;p) \| \le 
\G|p|+\sum_{k=2}^\io \G^{2k}\,
\fra{|p|^{2k-1}}{(2k-1)!}\, e^{c_{N-1} |p|} \,\le\,
\G|p| e^{d_N |p|} , \Eq(5.6)$$
where $\G$ is a suitable constant derived in \cita{GG1}
and $d_N=c_{N-1}+\G$.

\*
\0{\bf Remark.} The step leading from \equ(5.5) to \equ(5.6) is
non-trivial and the possibility of bounding the small divisors
$x_{\ell}^2$ and the sum over self--energy clusters, after defining the
propagators as above, is the main technical aspect of the work in
\cita{GG2}. We take the existence of $\G$ from \cita{GG2}.
We do not repeat here the analysis performed in
Sects. 5 and 6 (and the corresponding Appendices) of \cita{GG2}:
the constant $\G^2$ has been called
$\lis\e^{-1}$ in Theorem 1 of \cita{GG2}.
\*

Thus the inductive assumption holds for all $N$,
the constants $c_N,d_N$ can be taken $c 2^N$ for some $c$,
and for all $x$ on scale $N$ one has
$$ \| (g^{[N]})_B(x;p) \| \le \fra{|p|}{x^2}e^{2^N c \,|p|} .
\Eq(5.7) $$
This leads to a bound on $(\V h^{(N)})_B(\Bps,p)$,
via \equ(3.2) and \equ(3.3):
$$ \eqalign{ | (\V h^{(N)})_B(\Bps,p) | &\le 
\sum_{k=1}^\io\sum_\Bn
\sum_{\Th_{\Bn,\g}}
\fra1{|\L(\th)|!}\Big(
\mathop{{\prod}^{*}}_{\ell\in\L(\th)\atop n_\ell\ge 0} 
\fra1{x_{\ell}^2}|p|
e^{c_{n_{\ell}} |p|}\Big)\Big(\prod_{\ell\in\L(T)\atop n_\ell=-1} 
||g_\ell^{[-1]}||\Big)
\Big(\prod_{v\in V(T)} || F_{v}|| \Big)\le\cr
&\le \sum_{k=1}^\io \G^{2k} \fra{|p|^{2k-1}}{(2k-1)!}
e^ {c 2^{n_{\max}}\,|p|}
\le \sum_{k=1}^\io \G^{2k} \fra{|p|^{2k-1}}{(2k-1)!}
e^ {c' k\,|p|} \le \G^2 |p|e^{ \G |p|e^{c'|p|}} ,\cr}
\Eq(5.8) $$
because the maximum scale $n_{\max}$ of the lines of a graph with $k$
lines can be at most $\log_2 (bk)$ by our assumption that $f$ is a
trigonometric polynomial: in fact the maximum momentum on a line can be
$|\Bn|\le b_0 k$ (for $f$ a trigonometric polynomial), so that the
smallest $x$ can be $\fra{C_0}{b_0 k}$ if $C_0$ and $\t=1$ are the
Diophantine constants, hence the scale of $x$ can be at most $\log_2(
b_0 k/4)$ and $c 2^{n_{\max}}\le c' k$ for a suitable $c'$.

Therefore the functions $(\V h^{(N)})_B(\Bps;p)$ are entire
and bounded by
$$ |(\hh^{(N)})_B(\Bps;p)|\le \G^2 |p| e^{ \G |p|e^{c'|p|}} 
\Eq(5.9) $$
{\it independently of $N$}. Since for $|p|$ small the functions
$(\hh^{(N)})_B(\Bps;p)$ converge to $\V F(p)$, simply because of 
the bound $|\fra1{k!}\dpr^k\hh(\Bps,0)|\le D C^k k!$ on the Taylor 
coefficients at the origin of $\hh(\Bps,\h)$, it follows
that $\V F(\Bps;p)$ is entire in $p$ and everywhere the limit
of $(\V h^{(N)})_B(\Bps;p)$.
Therefore (by Vitali's theorem) $\V F(\Bps;p)$ is not only
holomorphic in $p$ near $0$ (which is a result on the Lindstedt series
at the origin) but it is entire. In particular it is holomorphic
around the real axis and therefore by \equ(4.4) $\hh(\Bps;\h)$
is Borel summable.

\*\*
\section(6, Concluding remarks)

(1) It is interesting to stress that the recursive bounds on the
Borel transform of the propagators in \equ(5.4) have been derived without 
making use of the cancellations that played such an essential role in the 
theory in \cita{GG2} (and \cita{GG1}) and by ``undoing'' at each step the 
resummations which led to the construction of $\hh$ and to the proof of 
Proposition 1, see \equ(5.3) above. 
However the properties of $\hh$ and the result of Proposition 1
have played a key role in the derivation of \equ(4.3) and \equ(4.4). 
Without the uniform bounds (in $N$)
on the convergence radii in $p$ of the series expressing
$(\hh^{(N)})_B$, which depend on the cancellations and on the resummations,
the bounds in Sect. \sec(5) or, in the non-trigonometric case,
of Appendix \sec(A1), would remain the same
but they would be useless for our purposes of establishing
\equ(4.4) and the Borel summability.

(2) The assumption that $f$ is a trigonometric polynomial has been
heavily exploited and an extra idea is necessary to deal with the more
general case of analytic $f$: this is discussed in Appendix \sec(A1).

(3) It has been remarked above that the resummation procedure is
based on several arbitrary \ap choices which therefore may lead to
the existence of several solutions of $\V h$ with the same asymptotic
series at $\e=0$.  All the choices in \cita{GG2}, as well as that
used in \cita{GG1} (see Appendix \sec(A2)),
lead however to a Borel summable series: this proves 
that all solutions coincide and the results of the resummations are
independent of the particular choices {\it provided} the Diophantine
constant $\t$ is $\t=1$ (hence $r=2$ and $\Bo$ is a Diophantine
vector with $\t=1$, \eg $\o_1/\o_2$ is a quadratic irrational).

(4) If $\t>1$, hence if $r>2$, the problem of Borel summability and of
independence of the result from the summation method remains open.

(5) The existence or non-existence of solutions $\V h$ which are
$C^{\io}$ at the origin and give solutions to the equations of
motion but which are not Borel summable is also an open
problem. Note that {\it even in the case of non-resonant motions}
the problem of the uniqueness at fixed $\Bo$ is not trivial,
and also the recent results in \cita{BT} do not exclude
the possibility of other quasi-periodic motions besides those
constructed through the KAM algorithm.

(6) The case $\e<0$ with $\Bb_0$ a minimum point for $f_{\V0}(\Bb)$,
\ie of elliptic motions is quite different. We have used on
purpose the resummation technique of \cita{GG2} which works for
hyperbolic as well as for elliptic resonances, but the present
results still only apply to the hyperbolic case and it is not
clear whether the above techniques can be extended to prove Borel
summability of the parametric equations for $\V h$ in elliptic cases.

\*\*
\appendix(A1, Analytic, non-polynomial perturbations)

\0In this Appendix we want to prove real analyticity of $\V F(\Bps,p)$
in the case $f$ is a generic analytic function 
(rather than a trigonometric polynomial);
in this case the function $\V F(\Bps,p)$ will turn out not be entire but 
only analytic in a strip of width $2 \k>0$ around the real axis. 

The proof of this claim will be based on an inductive assumption on 
$(g^{[n]})_{B}(x;p)$ formulated by introducing the matrix $\wt g^{[n]}(x;\h) =
\fra{\h^2}{x^2+\MM^{[\le n]}(x;\h)}$ for {\it all} $|x| < \g_{n-1}$.
Note that if $\chi_n(x)\defi \openone(\g_n\le|x|<\g_{n-1})$
is the indicator of the scale of $x$, the propagator $g^{[n]}(x;\h)$
is given by $g^{[n]}(x;\h)= \chi_n(x)\wt g^{[n]}(x;\h)$.
Furthermore the matrices $g^{[n]}(x;\h)$ satisfy the recursive equations
$$ \big(\wt g^{[n]}(x;\h) \big)^{-1} =
\big(\wt g^{[n-1]}(x;\h) \big)^{-1} + \h^{-2} \MM^{[n]}(x;\h) ,
\qquad \forall |x| < \g_{n-1}\;.\Eqa(A1.2) $$
We suppose, inductively, that for $\k_0=\sqrt{\|M_0\|}$ one has
$$ \| (\wt g^{[n]})_B(x;p) \|\,\le\, K_0\fra{|p|}{x^2}\, 
e^{(c_n +c'_{n}|x|^{-1/2})|p|+\k_0|\Im p||x|^{-1}}\;,\qquad\forall |x| <
\g_{n-1}\;,\qquad n\ge 0\;.
\Eqa(A1.3) $$
Note that for $n=0$ the explicit expression of $(\wt g^{[0]})_B$ given
by \equ(5.2) implies that \equ(A1.3) is valid with $K_0=\sqrt{s}$ (where 
$s$ is the dimension of the non trivial block in $M_0$)
and $c_0=c_0'=0$. The constant 
$K_0$ comes from our choice of the matrix norm $\|M\|=\max_j\sum_i|M_{ij}|$:
with this choice for any $d\times d$ matrix we have $d^{-1/2}
\|M\|_2\le\|M\|\le d^{1/2}\|M\|_2$ where 
$\|\cdot\|_2$ is the spectral norm, so that in particular $\|\sin M/M\|,
\|\cos M\|\le d^{1/2}$.

Assuming \equ(A1.3) for $n\le N-1$ and performing all convolutions
along the straight line from the origin to $p$, we get the following bound on
$(\MM^{[N]})_{B}(x;p)$:
$$ \eqalign{&
\big\|\big(\h^{-2} \MM^{[N]}\big)_B(x;p)\big\|\le\cr
&\le \sum_{k=2}^\io \sum_{T\in\SS^\RR_{k,N-1}}\fra1{|\L(T)|!}
\Big(\mathop{{\prod}^{*}}_{\ell\in\L(T)\atop n_\ell\ge 0}
K_0\fra{|p|}{x_\ell^2}e^{
(c_{n_\ell}+c'_{n_\ell} \g_{n_\ell}^{-1/2})|p| +
\k_0|\Im p|\g_{n_\ell}^{-1}}\Big) \cr
& \qquad \qquad \Big(\prod_{\ell\in\L(T)\atop n_\ell=-1}\|
g_\ell^{[-1]}\|\Big)\Big(\prod_{v\in V(T)}\|F_v\|\Big)
\cr
&\le \sum_{k=2}^\io 
\G^{2k} \fra{|p|^{2k-3}}{(2k-3)!} e^{(c_{N-1}+c'_{N-1} \g_{N-1}^{-1/2})|p| +
\k_0|\Im p|\g_{N-1}^{-1}}\,e^{-2\k 2^N}\,\le 
D_0|p|\,e^{d_N|p|}\,e^{-\k 2^N} ,\cr}
\Eqa(A1.4) $$
where $\G$ and $\k$ are deducible from Appendix A3 of \cita{GG2}, 
$D_0=\G^3$, $d_N=c_{N-1}+c_{N-1}'\g_{N-1}^{-1/2}$ and,
using that $\g_{N-1}^{-1}\le 4C_0^{-1}2^N$, we chose $|\Im p|$ so small that
$4\k_0 C_0^{-1}|\Im p|\le \k$. 

Using \equ(A1.4) and \equ(A1.2) we can get a bound
on $(\wt g^{[N]})_{B}(x;p)$:
$$\eqalign{
& \big\|\big(\wt g_N\big)_B(x;p)\big\|\le 
\Big(K_0\fra{|p|}{x^2} e^{(c_{N-1}+c'_{N-1}|x|^{-1/2})|p|+
\k_0|\Im p||x|^{-1}}\Big)* \cr
& * \sum_{k=0}^\io \Big[
\Big(K_0\fra{|p|}{x^2} e^{(c_{N-1}+c'_{N-1}|x|^{-1/2})|p|+
\k_0|\Im p||x|^{-1}}\Big)*\Big(D_0 |p|e^{d_N|p|}e^{-\k 2^N}\Big)\Big]^{*k} , 
\cr}\Eqa(A1.5)$$
and, since $|p|*(|p|*|p|)^{*k}=\fra{|p|^{4k+1}}{(4k+1)!}$ for $k\ge0$,
the $k$-th term in the sum is bounded by
$$K_0\fra{(K_0D_0e^{-\k 2^N})^k}{(4k+1)!}\fra{|p|}{x^2} \fra{|p|^{4k}}
{x^{2k}} e^{(d_N+ c'_{N-1}|x|^{-1/2})|p|+
\k_0|\Im p||x|^{-1}} \;.\Eqa(A1.6)$$
Summing \equ(A1.6) over $k\ge 0$ and comparing the result with the inductive
assumption \equ(A1.3), we see that we can take $c_N=d_N=c_{N-1}+c_{N-1}'
\g_{N-1}^{-\a}$ and $c_N'=c_{N-1}'+K_0D_0e^{-\k 2^N}$.
Solving the iterative equations for $c_N,c_N'$, we see that,
for $x$ on scale $n$, $(\wt g^{[n]})_B(x;p)$ can be bounded as
$$ \| (\wt g^{[n]})_B(x;p) \|\,\le\, K_0\fra{|p|}{x^2}\, 
e^{c 2^{n/2}|p|+\k_1|\Im p|2^n}\;,\qquad
\g_n\le|x|<\g_{n-1}\;,\qquad n\ge 0\;,
\Eqa(A1.6a) $$
for some constants $c,\k_1$.

Plugging this bound into the expansion for $\hh_B^{(N)}(\Bps,p)$,
denoting by $N(\th)$ the maximal scale in $\th$ and choosing
$|\Im p|$ small enough, we finally get
$$ \eqalign{
& | (\V h^{(N)})_B(\Bps,p) | \le\cr 
& \quad \le\sum_{k=1}^\io\sum_\Bn\sum_{n_0\ge 0}
\sum_{\th\in\Th_{\Bn,\g}\atop N(\th)=n_0}
\fra1{|\L(\th)|!} \Big(\mathop{{\prod}^{*}}_{\ell\in\L(\th)
\atop n_\ell\ge 0} \fra{K_0}{x_{\ell}^2}|p|
e^{c 2^{n_{\ell}/2} |p|+\k_1|\Im p|2^{n_\ell}} \Big) \cr
& \qquad \qquad \Big(\prod_{\ell\in\L(T)\atop n_\ell=-1} 
\|g_\ell^{[-1]}\|\Big) \Big(\prod_{v\in V(T)} \| F_{v}\| \Big) \le \cr
& \quad \le \sum_{k=1}^\io \sum_{n_0\ge 0}
\G^{2k} \fra{|p|^{2k-1}}{(2k-1)!} e^ {c 2^{n_0/2}\,|p|+\k_1|\Im p|2^{n_0}}
e^{-2\k 2^{n_0}}\le \G' |p|e^{ c'|p|^2} ,\cr}
\Eqa(A1.7)$$
for some new constants $\G',c'$.
Performing the limit $N\to\io$ in \equ(A1.7), we see that 
$\V F(\Bps,p)$ satisfies the same bound
$$ | \V F(\Bps,p) | \le \G'|p|\, e^{c'|p|^2}, \qquad{\rm for}\ |\Im p|\le \s ,
\Eqa(A1.8) $$
for some $\s$ small enough, so that $\V F(\Bps,p)$ is analytic (in $p$) 
in a strip and
therefore $\V h(\Bps,\h)$ is Borel summable (in $\h$).

\*\*
\appendix(A2,Comparison with the method of \cita{GG1})

\0In this Appendix we briefly discuss how the function $\hh$
constructed in \cita{GG1} can be identified with the Borel summable
function of Proposition 1. By the uniqueness in the class of
Borel summable functions, it is enough to prove that also
the function $\hh$ of \cita{GG1} is Borel summable.

We begin by reviewing the differences of the construction
envisaged in \cita{GG1} with respect to that of \cita{GG2}.
Trees, labels and clusters are defined in the same way
as in items (a) to (h) of Section \sec(3). What changes
is the definition of the propagators, which is iterative.
By writing $x=\Bo\cdot\Bn_{\ell}$, $\Bn_{\ell}
\neq\V0$, we set $g^{[0]}_{\ell}=1/x^{2}$, and, for $k \ge 1$,
$$ g^{[k]}_{\ell}=\fra{\h^2}{x^2+M^{[k]}(x;\h)} .
\Eqa(A2.1) $$
with $M^{[k]}(x;\h)$ defined as
$$ M^{[k]}(x;\h) = \sum_{T} \VV_{T}(x;\h) ,
\Eqa(A2.2) $$
where the sum is restricted to the self-energy clusters $T$
with scale $n_{T} \ge n+3$, where $n$ is such that $\g_{n-1} \le |x| < \g_{n}$,
and the self-energy value is given by
$$ \VV_{T}(\Bo\cdot\Bn;\h) = -{\h^2\over |\L(T)|!}
\Big(\prod_{v \in V(T)} F_{v} \Big)
\Big(\prod_{\ell \in \L(T)} g^{[k-1]}_{\ell} \Big) .
\Eqa(A2.3) $$
Note that $k$ labels the iterative step and, in principle,
it has no relation with the scale $n$ of $x$. However
the matrices $M^{[k]}(x;\h)$ are obtained from resummations
of self--energy clusters with height up to $k=n$
(for definitions and details we refer to \cita{GG1},
where the self--energy clusters are called self--energy graphs).
Hence they stop flowing at $k=n$ if $x$ is on scale $n$:
this means that $M^{[k]}(x;\h)=M^{[n]}(x;\h)$ as soon as $k \ge n$.

Then $\hh^{(N)}$ will be expressed in terms of trees as
in \equ(3.3), if $\Val(\theta)$ is defined as in \equ(3.2),
with $g^{[n_{\ell}]}_{\ell}$ replaced with $g^{[N]}_{\ell}$,
and $\hh^{(N)}$ is obtained as the limit of $\hh^{(N)}$ as $N\to\io$.

Let us consider for simplicity the case of polynomial perturbations.
Then we can proceed as in Sec. \sec(5), and prove by induction the bound
$$ \big\| \big(M^{[k]} \big)_B(x;p)\big\|\le \G|p|
e^{ d_k\, |p|}, \qquad {\rm for\ all}\ x ,
\Eqa(A2.4) $$
for all $p$ complex. Supposing inductively the bound \equ(A2.4) we
obtain that the Borel transform of $g^{[k]}_{\ell}$ can be bounded as
$$ \|(g^{[k]})_B(x;p) \| \le \fra{|p|}{x^2} e^{c_k|p|} .
\Eqa(A2.5) $$
This is trivial for $k=0$, as $(g^{[0]})_{B}(x;p)=p/x^2$,
while it follows from \equ(A2.4) for $n\ge1$ by the
inductive hypothesis (and it can be proved as the analogous
bound in Sec. \sec(5)). Therefore we can write
$M^{[N]}(x;\h)$ according to \equ(A2.2), and its
Borel transform  $(M^{[N]})_{B}(x;p)$ can be computed
and bounded as done in \equ(5.5) and \equ(5.7), so that at the end
the bound \equ(A2.4) is obtained for $k=N$. In particular
the bounds \equ(A2.5) on the propagators imply
that $\hh^{(N)}$ admits the same bound \equ(5.9) found
in Sec. \sec(5). Therefore we can take the limit for $N\to\io$,
and Borel summability for $\hh$ follows.

\*


\rife{BT}{1}{
H. Broer, F. Takens,
{\it Unicity of KAM tori},
Preprint, Groningen, 2005.}
\rife{GBG}{2}{
G. Gallavotti, F. Bonetto, G. Gentile,
{\it Aspects of ergodic, qualitative and statistical theory of motion},
Texts and Monographs in Physics,
Springer, Berlin, 2004. }
\rife{GG}{3}{
G. Gallavotti, G. Gentile,
{\it Majorant series convergence for twistless KAM tori},
Ergodic Theory and Dynamical Systems
{\bf 15} (1995), 857--869. }
\rife{GG1}{4}{
G. Gallavotti, G. Gentile,
{\it Hyperbolic low-dimensional invariant tori and summations
of divergent series},
Communications in Mathematical Physics
{\bf 227} (2002), no. 3, 421--460. }
\rife{GG2}{5}{
G. Gallavotti, G. Gentile,
{\it Degenerate elliptic resonances},
Communications in Mathematical Physics
{\bf 257} (2005), 319--362. }
\rife{JLZ}{6}{
\`A. Jorba, R. de la Llave, M. Zou,
{\it Lindstedt series for lower-dimensional tori},
``Hamiltonian systems with three or more degrees of freedom''
(S'Agar\'o, 1995), 151--167,
NATO Adv. Sci. Inst. Ser. C Math. Phys. Sci., 533,
Kluwer Acad. Publ., Dordrecht, 1999. }
\rife{Ne}{7}{
F. Nevanlinna,
{\it Zur Theorie der Asymptotischen Potenzreihen},
 Annales Academiae Scientiarum Fennicae. Series A I. Mathematica
{\bf 12} (1916), 1--18.} 
\rife{So}{8}{
A.D. Sokal, 
{\it An improvement of Watson's theorem on Borel summability},  
Journal of Mathemnatical Physics
{\bf 21} (1980), no. 2, 261--263. }

\biblio 

\ciao